\documentclass[12pt,preprint]{article}
\usepackage{amssymb}
\usepackage{amsmath}
\usepackage{graphics}

\newcommand{\eqb}{\begin{equation}}
\newcommand{\eqe}{\end{equation}}
\newcommand{\dmb}{\begin{displaymath}}
\newcommand{\dme}{\end{displaymath}}

\newcommand{\eab}{\begin{eqnarray}}
\newcommand{\eae}{\end{eqnarray}}

\newcommand{\be}{\begin{equation}}
\newcommand{\ee}{\end{equation}}

\begin{document}

\begin{titlepage}
\begin{flushright} 
\end{flushright}
\vspace{0.6cm}

\begin{center}
\Large{Potential for inert adjoint scalar field in SU(2) Yang-Mills thermodynamics}
\vspace{1.5cm}

\large{Francesco Giacosa$^\dagger$ and Ralf Hofmann$^*$}

\end{center}
\vspace{1.5cm} 

\begin{center}
{\em $\mbox{}^\dagger$ Institut f\"ur Theoretische Physik\\ 
Universit\"at Frankfurt\\ 
Johann Wolfgang Goethe - Universit\"at\\ 
Max von Laue--Str. 1\\ 
60438 Frankfurt, Germany}
\end{center}
\vspace{1.5cm}

\begin{center}
{\em $\mbox{}^*$ Institut f\"ur Theoretische Physik\\ 
Universit\"at Heidelberg\\ 
Philosophenweg 16\\ 
69120 Heidelberg, Germany}
\end{center}
\vspace{1.5cm}

\begin{abstract}

A scalar adjoint field is introduced as a spatial average over (anti)calorons
in a thermalized SU(2) Yang-Mills theory. 
This field is associated with the thermal ground state in the deconfining phase and
acts as a background for gauge fields of trivial topology.
Without invoking detailed microscopic information we study the properties of
 the corresponding potential, and we discuss its thermodynamical implications.
We also investigate the gluon condensate at finite temperature relating 
it to the adjoint scalar field.

\end{abstract} 

\end{titlepage}

\section{Introduction}

The potential importance of topological field configurations in generating a
finite correlation length in the dynamics of thermalized, nonabelian gauge
fields \cite{Linde1980} was emphasized a long time ago \cite{Polyakov}. In
particular, calorons, i.e. instantons at finite temperature $%
T, $ play an important role in the description of pure SU(2) and SU(3) Yang-Mills
theories in their deconfining phase. However, the inclusion of topologically 
nontrivial field configurations when evaluating thermodynamical quantities 
is complicated because of the nonlinearity of the theory \cite{MP}. Also, there is no a priori 
infrared-cutoff when integrating out residual quantum fluctuations 
about these finite-action configurations \cite{Diakonov2004}.

In \cite{Hofmann} the possibility that a scalar adjoint field $\phi $ can
act as a thermal average upon calorons and anticalorons was addressed. 
More precisely, BPS saturated and stable (trivial holonomy)
configurations of topological charge modulus $|Q|=1$, Harrington-Shepard
solutions \cite{HS1977}, were integrated into an adjoint scalar field $\phi $
in the deconfining phase. This field $\phi$, which is part of the thermal ground state, 
affects thermodynamical quantities. The scalar field $\phi$
induces an adjoint Higgs mechanism: some of the gauge modes acquire a
(temperature-dependent) mass on tree-level. Notice that rotational
symmetry in the thermal system together with the perturbative 
renormalizability of the fundamental Yang-Mills action 
\cite{'t HooftVeltmann} forbid the emergence of any 
other composite field, induced by BPS saturated, nonpropagating,
fundamental field configurations, but an adjoint scalar
field\footnote{For a more detailed discussion see \cite{Hofmann}. The 
argument involves the fact that perturbative renormalizability implies
that $\phi$ transforms homogeneously under changes of the gauge. On the
fundamental level the only homogeneously transforming quantity is the
field strength and (nonlocal) products thereof which can always be
reduced to spin-0 and spin-1 representations of SU(2). While
the former is irrelevant on BPS saturated configurations the 
latter plays an important role. 
A rigorous version of the here-sketched argument will be presented 
in \cite{Ralf2007}.}.     
 
To be more definite, let us denote by $\mathcal{L}_{YM}=-\frac{1}{4}(F_{\mu \nu
}^{a})^{2}$ the fundamental $SU(2)$ Yang-Mills Lagrangian, with $F_{\mu \nu
}^{a}=\partial _{\mu }A_{\nu }^{a}-\partial _{\nu }A_{\mu
}^{a}-gf^{abc}A_{\mu }^{b}A_{\nu }^{c}$ ($g$ is the fundamental coupling).
In the partition function all possible configurations of the field $A_{\mu }$
have to be considered, including the topologically nontrivial ones. Upon the emergence 
of the scalar adjoint field $\phi$ we are led to consider the
effective Lagrangian $\mathcal{L}_{{\tiny \mbox{eff}}}=\mathrm{Tr}\left( 
\frac{1}{2}\,G_{\mu \nu }^{E}G_{E}^{\mu \nu }+\left( D_{\mu }\phi \right)
^{2}+V(|\phi |^{2})\right) $ where $G_{\mu \nu }^{E,a}=\partial _{\mu
}a_{\nu }^{a}-\partial _{\nu }a_{\mu }^{a}-ef^{abc}a_{\mu }^{b}a_{\nu }^{c}.$
The fields $a_{\mu }^{a}$ refer to (coarse-grained) topologically trivial fluctuations 
and $e$ denotes the {\sl effective} coupling constant. The effect of
topology is embodied in the scalar adjoint field $\phi ,$ which acts as
a background for the dynamics of propagating gauge fields. While the field $\phi $ acts as an 
infrared-cutoff in the spatial coarse-graining performed in the fundamental theory 
\cite{Hofmann} it represents 
an ultraviolet cutoff in the coarse-grained theory, implying a rapidly
converging loop expansion \cite{Hofmann,Hofmann2006Sept}. 
The potential $V(|\phi |^{2})=\Lambda ^{6}/|\phi |^{2}$, where $\Lambda $ is the
Yang-Mills scale, was evaluated in \cite{Hofmann} by performing the spatial average over 
calorons explicitely.

In this work we would like to show that apart from general properties such as 
gauge invariance and the periodicity of $\phi$ in the euclidean time 
the potential $V(|\phi |^{2})$
follows by requiring BPS saturation. That is, no {\sl explicit} calculation 
involving calorons is needed to derive $V(|\phi |^{2})$. Over and above the existence of a Yang-Mills
scale $\Lambda$, which needed to be assumed in \cite{Hofmann}, turns out to be redundant 
since $\Lambda$ is shown to be a purely 
nonperturbative constant of integration. 

The article is organized as follows: In Sec.\,\ref{GI} we discuss the implications of 
BPS saturation for the dynamics of the field $\phi$. The properties of (coarse-grained) 
topologically trivial fluctuations and the combined 
effect of the ground state and the excitations on basic thermodynamical quantities 
are investigated in Sec.\,\ref{q=0}. In Sec.\,\ref{gc} we show that a 
gluon condensate emerges in the framework of the effective theory and that its 
$T$-dependence is in agreement with lattice results \cite{millerrev,miller,Langfeld}. 

\section{Gauge invariance, BPS saturation, and inertness\label{GI}}

In the deconfining phase of thermalized SU(2) Yang-Mills theory 
we study the possibility that an adjoint scalar field $\phi $
describes (part of) the thermal ground state. Namely, we postulate that $%
\phi =$ $\phi (\tau ),$ where $0\leq \tau \leq \beta \equiv 1/T$, emerges
from by virtue of a {\sl spatial} average over (anti)selfdual fundamental field
configurations in euclidean spacetime. Intuitively, we assume that the
nontrivial nature of the Yang-Mills ground state can be described 
by a scalar-adjoint field $\phi$.

Without the need to perform the average explicitely the field $\phi $ enjoys
the following general properties as a consequence: (i) Since $\phi $ is
obtained by a spatial coarse-graining over noninteracting, \textsl{stable},
BPS saturated field configurations (topology changing energy and pressure
free fluctuations) it is itself BPS saturated, thus the associated energy
density vanishes. (ii) Originating from periodic-in-$\tau $ field
configurations (in a given gauge) it is itself periodic. (iii) The gauge
invariant modulus $|\phi |$ does not depend on spacetime (trivial expansion
into Matsubara frequencies due to coarse-graining over energy and pressure
free configurations).

We will now show that conditions (i)-(iii) uniquely fix the potential $V$ for the
field $\phi\equiv \phi^a(\tau)\,\lambda^a$ ($\mbox{tr}\lambda_a
\lambda_b=2\delta_{ab},\,a=1,2,3$) when working with a canonical
kinetic term in its euclidean Lagrangian density $\mathcal{L}_{\phi}$ 
\footnote{%
As long as this term contains two powers of time derivatives this is not a
constraint on generality due to (iii). Moreover, although we ignore the
connection to the microscopic physics in the present work we surely can make
an appropriate choice of gauge such that the coarse-graining over \textsl{%
noninteracting} topological defects generates $A_{\mu }=0$ on the
macroscopic level.}: 
\begin{equation}  \label{lagphi}
\mathcal{L}_{\phi }=\mbox{tr}\,\left( \left(\partial_{\tau }\phi \right)
^{2}+V(\phi ^{2})\right)
\end{equation}
Since the coarse-graining is over exact solutions to the Yang-Mills
equations the emerging field $\phi $ must minimize the effective action.
Thus $\phi$ satisfies the Euler-Lagrange equations subject to Eq.\,(\ref%
{lagphi}): 
\begin{equation}
\partial _{\tau }^{2}\phi ^{a}=\frac{\partial V(\left\vert \phi \right\vert
^{2})}{\partial \left\vert \phi \right\vert ^{2}}\phi ^{a}\leftrightarrow
\partial _{\tau }^{2}\phi =\frac{\partial V(\phi ^{2})}{\partial \phi ^{2}}%
\phi\,,  \label{orbiteq}
\end{equation}%
where $|\phi|\equiv \sqrt{\frac{1}{2}\,\mbox{tr}\,\phi^2}$. The gauge
invariance of the potential $V=V(|\phi|^2)$ (central potential) in Eq.\,(\ref%
{orbiteq}) implies that the solution has to describe motion in a plane of
the three-dimensional vector space spanned by the Lie-algebra valued
generators of SU(2) Yang-Mills theory. (The angular momentum is a
constant of motion in a central potential.) 

Without restriction of generality (a global gauge choice) we choose the
plane $(\phi ^{1},\phi ^{2},0)$. Thus the solution takes the following form: 
\begin{equation}
\phi =\left\vert \phi \right\vert \lambda _{1}\exp (i\lambda _{3}\theta
(\tau ))=\left\vert \phi \right\vert \left( \lambda _{1}\cos (\theta (\tau
))+\lambda _{2}\sin (\theta (\tau ))\right)\,,  \label{phi1}
\end{equation}
or in components 
\begin{equation}
(\phi ^{1},\phi ^{2},\phi ^{3})=\left\vert \phi \right\vert (\cos(\theta
(\tau)),\,\sin(\theta(\tau )),\,0)\,.  \label{phi2}
\end{equation}
According to (ii) the function $\theta(\tau)$ needs to satisfy the following
condition 
\begin{equation}  \label{theta}
\theta(\tau+\beta)=\theta(\tau)+2\pi n\,,
\end{equation}
where $n$ is an integer. Finally, condition (i) implies the vanishing of the
(euclidean) energy density $\mathcal{H}_{E}(\phi)$: 
\begin{equation}
\mathcal{H}_{E}(\phi)=\mbox{tr}\,\Big(\left( \partial _{\tau }\phi \right)
^{2}-V(\phi ^{2})\Big) =2\left( (\partial _{\tau }\phi
^{a})^{2}-V(\left\vert \phi \right\vert ^{2})\right) =0\text{ }\,,\ \ \
\forall \beta\,.  \label{bpsphi}
\end{equation}

\noindent Substituting Eq.\,(\ref{phi1}) into Eq.\,(\ref{bpsphi}) we have 
\begin{equation}
\left\vert \phi \right\vert ^{2}\left(\partial _{\tau }\theta (\tau )\right)
^{2}-V(\left\vert \phi \right\vert ^{2})=0\,.  \label{uno}
\end{equation}%
According to (iii) the potential $V(\left\vert \phi \right\vert ^{2})$ does
not depend on $\tau$. As a consequence of Eq.\,(\ref{uno}), we then have $%
\partial _{\tau }\theta (\tau )=\mbox{const}$. Together with Eq.\,(\ref%
{theta}) this yields: 
\begin{equation}
\theta (\tau )=\frac{2\pi }{\beta }n\tau\,  \label{thetan}
\end{equation}%
up to an inessential constant phase (global gauge choice). Notice that the
case $n=0$ is excluded if we impose that\footnote{%
In a similar way, the case $Q=0$ is excluded for BPS saturated, microscopic
field configurations if we insist on a nonvanishing action.} $V\not=0$. Now
Eq.\,(\ref{thetan}) implies that $\partial _{\tau }^{2}\theta (\tau )=0$,
and thus we obtain from Eqs.\, (\ref{phi1}) and (\ref{orbiteq}) that 
\begin{equation}  \label{due}
\left( \partial _{\tau }\theta (\tau )\right) ^{2}=-\frac{\partial
V(\left\vert \phi \right\vert ^{2})}{\partial \left\vert \phi \right\vert
^{2}}\,.
\end{equation}
Eliminating $\left( \partial _{\tau }\theta (\tau )\right) ^{2}$ from Eqs.\,(%
\ref{uno}) and (\ref{due}) we have: 
\begin{equation}
\frac{V(\left\vert \phi \right\vert ^{2})}{\left\vert \phi \right\vert ^{2}}%
=-\frac{\partial V(\left\vert \phi \right\vert ^{2})}{\partial \left\vert
\phi \right\vert ^{2}}\,.  \label{diffeq}
\end{equation}%
%
%
%
%
%
%
Notice that Eq.\,(\ref{diffeq}) is valid for all values of $\beta$. The
unique solution to the first-order differential equation (\ref{diffeq}) is 
\begin{equation}
V(\left\vert \phi \right\vert ^{2})=\frac{\Lambda ^{6}}{\left\vert \phi
\right\vert ^{2}}  \label{potential}
\end{equation}%
where the mass scale $\Lambda$ enters as a constant of integration. On
dimensional grounds $\Lambda$ has to appear with the sixth power. We
interprete $\Lambda$ as the Yang-Mills scale which, however, is not
operational on the level of BPS saturated dynamics, see below. 
(On this level the energy-momentum tensor vanishes.)

Let us now determine the modulus $\left\vert \phi \right\vert .$ By
inserting the potential $V(\left\vert \phi \right\vert ^{2})=\Lambda
^{6}/\left\vert \phi \right\vert ^{2}$ and Eq.\,(\ref{thetan}) into Eq.\,(%
\ref{uno}) we have 
\begin{equation}
\left\vert \phi \right\vert =\sqrt{\frac{\Lambda ^{3}}{2\pi |n|\,T}}\,.
\end{equation}%
This implies that the field $\phi$ vanishes in a power-like way with
increasing temperature. The value of the integer $n$ can not be determined
within the macroscopic approach we have applied to deduce $\phi$'s
potential. Microscopically, one observes that the definition of $\phi$'s
phase does only allow for the contribution of Harrington-Shepard solutions 
\cite{HS1977} of topological charge modulus $|Q|=1$ which implies that $%
n=\pm 1$ \cite{Hofmann}.

Finally, we point out the inertness of the field $\phi$. According to Eqs.\,(%
\ref{potential}) and (\ref{lagphi}) the square of the mass $M_\phi$ of
potential (radial) fluctuations $\delta\phi$ is given as (setting $|n|=1$) 
\begin{equation}  \label{massphi}
M_\phi^2=2\,\left.\frac{\partial^2 V}{\partial |\phi|^2}\right|_{|\phi|=%
\sqrt{\frac{\Lambda ^{3}}{2\pi\,T}}}=48\,\pi^2\,T^2\,.
\end{equation}
Thus $\frac{M_\phi^2}{T^2}=48\,\pi^2\gg 1$, and no thermal excitations
exist. On the other hand, we have $\frac{M_\phi^2}{|\phi|^2}=12\,\lambda^3$
where $\lambda\equiv\frac{2\pi T}{\Lambda}$. For $\lambda\gg 1$ one has that 
$\frac{M_\phi^2}{|\phi|^2}\gg 1$. In practice, $\lambda>\lambda_c=13.87$,
see \cite{Hofmann}. Since $|\phi|$ is the maximal resolving power allowed in
the effective theory we conclude that quantum fluctuations of the field $%
\phi $ do not exist.

\section{Topologically trivial fluctuations\label{q=0}}

\subsection{Effective Lagrangian and ground state}

For the reader's convenience we briefly repeat the derivation of \cite%
{Hofmann} leading to the complete ground-state description of SU(2)
Yang-Mills thermodynamics in its deconfining phase.

If topological fluctuations were absent then renormalizability \cite{'t
HooftVeltmann} would assure that the action of the fundamental theory is
form-invariant under the applied spatial coarse-graining. Since the
topological part is integrated into an inert field $\phi$ this still holds
true for the part of the effective action induced by $Q=0$-fluctuations $%
a_{\mu}$. We thus are confronted with the following, gauge invariant
effective Lagrangian for the dynamics of coarse-grained $Q=0$ fluctuations $%
a_\mu$ subject to the background $\phi$: 
\begin{equation}
\mathcal{L}_{{\tiny \mbox{eff}}}=\mathcal{L}\left[ a_{\mu }\right] =\mbox{tr}%
\,\left( \frac{1}{2}\,G^{E}_{\mu \nu }G_{E}^{\mu \nu }+\left( D_{\mu }\phi
\right) ^{2}+\frac{\Lambda ^{6}}{\phi ^{2}}\right)\,,\text{ }  \label{lageff}
\end{equation}%
%
%
%
%
%
%
where 
\begin{eqnarray}
G_{E}^{\mu \nu }&=&\partial _{\mu }a_{\nu }-\partial _{\nu }a_{\mu }-ie\left[
a_{\mu },a_{\nu }\right] =G_{E}^{a,\mu \nu }\frac{\lambda _{a}}{2}\,,  \notag
\\
a_{\mu }&=&a_{\mu }^{a}\frac{\lambda _{a}}{2}\,,\ \ \ \ D_{\mu
}\phi=\partial _{\mu}\phi -ie[a_{\mu },\phi ]\,,
\end{eqnarray}
and $e$ denotes an \textsl{effective} gauge coupling. According to Eq.\,(\ref%
{lageff}) the equation of motion for the field $a_\mu$ is: 
\begin{equation}
D_{\mu }G_{E}^{\mu \nu }=ie\left[ \phi ,D_{\nu }\phi \right]\,.  \label{ym1}
\end{equation}%
This is solved by the pure-gauge configuration $a_{\mu }=a_{\mu }^{gs}$
given as 
\begin{equation}
a_{\mu }^{gs}=\mp \delta _{\mu 4}\frac{2\pi }{\beta }\frac{\lambda _{3}}{2}=%
\frac{i}{e}\left( \partial _{\mu }\Omega \right) \Omega ^{\dagger }\text{
with }\Omega =e^{\pm i\frac{2\pi }{\beta }\tau \frac{\lambda _{3}}{2}%
}\Rightarrow D_{\nu }\phi =0
\end{equation}
The entire ground state thus is described by the $a_{\mu }^{gs},\phi$
implying a ground-state pressure $P^{gs}$ and energy density $\rho^{gs}$
given as $P^{gs}=-4\pi\Lambda^3 T=-\rho^{gs}$: The inclusion of gluon
fluctuations, which contribute to the dynamics of the ground-state, by
virtue of $a_{\mu }=a_{\mu }^{gs}$ after coarse-graining shifts the
vanishing results, obtained from BPS saturated configurations alone, to
finite values. This makes the Yang-Mills scale $\Lambda$ (gravitationally)
visible. Turning to propagating fluctuations $\delta a_\mu$ in the effective
theory it is advantageous to work in unitary gauge.

\subsection{Unitary gauge and Higgs mechanism}

By performing a gauge rotation\footnote{%
Notice that $U$ is smooth and antiperiodic. One can introduce a center jump
to make it periodic by sacrificing its smoothness \cite{Hofmann}. However,
the associated electric center flux does not carry any energy or pressure
and the periodicity of effective gluon fluctuations is maintained. These are
the physical reasons why the transformation to unitary gauge is admissible.} 
$U=e^{-i\frac{\pi }{4}\lambda _{2}}\Omega$ we have that $a_{\mu
}^{gs}\rightarrow Ua_{\mu }^{gs}U^{\dagger }-\frac{i}{e}\left( \partial
_{\mu }U\right) U^{\dagger }=0$ and $\phi =\lambda_{3}\left\vert \phi
\right\vert $. This is the unitary gauge. The field strength $G_{E}^{\mu
\nu} $ and the covariant derivative $D_{\mu }\phi$ are functionals of the
fluctuations $\delta a_\mu$ only. We have 
\begin{equation}
\mathcal{L}_{{\tiny \mbox{eff}}}^{u.g.}=\mathcal{L}\left[ \delta a_{\mu }%
\right] =\frac{1}{4}\left( G_{E}^{a,\mu \nu }[\delta a_{\mu }]\right)
^{2}+2e^{2}\left\vert \phi \right\vert ^{2}\left( \left( \delta a_{\mu
}^{(1)}\right) ^{2}+\left( \delta a_{\mu }^{(2)}\right) ^{2}\right) +2\frac{%
\Lambda ^{6}}{\left\vert \phi \right\vert ^{2}}\,.  \label{lagfirstord}
\end{equation}%
Fluctuations $\delta a_{\mu }^{(1,2)}$ are massive in a temperature
dependent way while the mode $\delta a_{\mu }^{(3)}$ remains massless
representing the fact that SU(2) is broken to its subgroup U(1) by the field 
$\phi$. One has 
\begin{equation}  \label{masssp}
m^2=m_{1}^2=m_{2}^2=4e^{2}\left\vert \phi \right\vert ^{2}\,,\ \ \ \
m_{3}^{2}=0\,.
\end{equation}

\subsection{Energy density, pressure and running coupling}

From the effective Lagrangian (\ref{lagfirstord}) we derive the energy
density $\rho$ and the pressure $p$ on the one-loop level. 
This is accurate on the 0.1\%-level as shown in 
\cite{Hofmann,KH2007}. This strong suppression of the effects of 
residual $Q=0$ quantum fluctuations in the effective theory takes place due 
to limited resolution, given by the modulus $|\phi|$, 
and due to emergent, temperature-dependent tree-level mass. 
Both phenomena introduce nonperturbative aspects into the loop expansion
based on the tree-level action Eq.\,(\ref{lagfirstord}) which render the
radiative corrections small.        

On the one-loop level we have
\begin{equation}
\rho =\rho_{3}+\rho _{1,2}+\rho _{gs}\,,\ \ \text{ }p=p_{3 }+p_{1,2}+p_{gs}
\,,  \label{rhop}
\end{equation}%
where the subscript 1,2 is understood as a sum over the two massive modes.
Explicitely, we have: 
\begin{eqnarray}
\rho_{3}&=&2\frac{\pi ^{2}}{30}\,T^{4},\text{ }\rho _{1,2}=6\int_{0}^{\infty
}\frac{k^{2}dk}{2\pi ^{2}}\frac{\sqrt{m^{2}+k^{2}}}{\exp (\frac{\sqrt{%
m^{2}+k^{2}}}{T})-1},\text{ }\rho _{gs}=2\frac{\Lambda ^{6}}{\left\vert \phi
\right\vert ^{2}}=4\pi \Lambda ^{3}T\,. \\
p_{3} &=&2\frac{\pi ^{2}}{90}T^{4},\text{ }p_{1,2}=-6\,T\int_{0}^{\infty }%
\frac{k^{2}dk}{2\pi ^{2}}\ln \left( 1-e^{-\frac{\sqrt{m^{2}+k^{2}}}{T}%
}\right)\,,\text{ }p_{gs}=-\rho _{gs}\,.
\end{eqnarray}
The effective coupling constant $e$ is a function of the temperature $e=e(T)$%
, and so is $m$. The function $e=e(T)$ is deduced by requiring the validity
of the Legendre transformation 
\begin{equation}
\rho =T\frac{dp}{dT}-p.  \label{Legendre}
\end{equation}%
in the effective theory.

By substituting the equations (\ref{rhop}) into (\ref{Legendre}) we obtain: 
\begin{equation}
4\pi \Lambda ^{3}=-6D(m)\frac{dm(T)}{dT},\text{ }D(m)=\int_{0}^{\infty }%
\frac{k^{2}dk}{2\pi ^{2}}\frac{m}{\sqrt{m^{2}+k^{2}}}\frac{1}{e^{\frac{\sqrt{%
m^{2}+k^{2}}}{T}}-1}\,.  \label{eqdiff}
\end{equation}
Solving the differential equation (\ref{eqdiff}) and inverting the solution,
the function $e(T)$ follows by virtue of Eq.\,(\ref{masssp}).

Eq.\thinspace (\ref{eqdiff}) is of first order. Thus a boundary condition
needs to be prescribed. It was shown in \cite{Hofmann} that the evolution at
low temperature decouples from the boundary physics at high temperature.
That is, there exists a low-temperature attractor to the evolution. This
attractor is characterized by a logarithmic pole, $e\sim -\log (T-T_{c})$
where $T_{c}=13.87\frac{\Lambda }{2\pi }$, signalizing the presence of a
phase transition, and by a plateau $e\equiv \sqrt{8}\pi$ for $T$ sufficiently larger
than $T_{c}$, indicating magnetic charge conservation for screened monopoles.
\begin{figure}[tbp]
\begin{center}
\leavevmode
\leavevmode
\vspace{5.5cm} \includegraphics{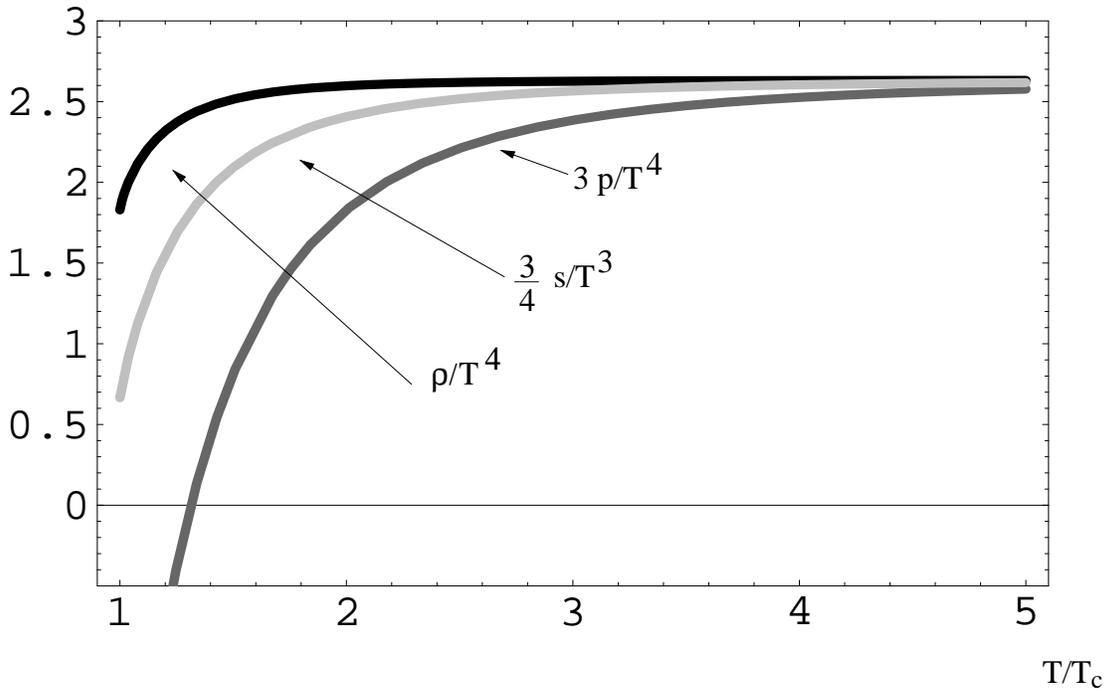}
\end{center}
\caption{Scaled energy density $\rho$ (black), pressure $p$ (dark gray), and entropy density $s$ 
(light gray) in the deconfining phase of SU(2) Yang-Mills thermodynamics.}
\label{Fig-1}
\end{figure}
In Fig.\,\ref{Fig-1} we indicate the (scaled) energy density $\rho$, pressure $p$, and entropy
density $s$ as functions of temperature. A detailed comparison of these results
with those obtained on the lattice (for both differential and integral
method) is carried out in the first reference of \cite{Hofmann}. While
there is good agreement for the infrared safe quantity entropy density 
$s$ the pressure $p$ becomes negative close to the phase boundary which 
is qualitatively in agreement with the result of the differential method but not 
with that of the integral method. 

\section{Gluon condensate\label{gc}}

We start with an intuitive discussion on the gluon condensate. As
emphasized in \cite{Diakonov:1995ea}, at $T=0$ instantons are responsible for a
nonzero gluon condensate \cite{Shifman:1978bx}. In fact, the average $%
\left\langle F_{\mu \nu }^{2}\right\rangle =-4\left\langle \mathcal{L}%
_{YM}\right\rangle =2\left\langle \overrightarrow{B}^{2}-\overrightarrow{E}%
^{2}\right\rangle $ is zero at any order in perturbation theory ($F_{\mu \nu
}$ is the fundamental stress-energy tensor). Instantons are selfdual 
solutions in euclidean spacetime, which, in Minkowski space, are interpreted as 
tunnelling events implying $E^{i,a}=\pm iB^{i,a}$ and so
generate a positive average $\left\langle F_{\mu \nu }^{2}\right\rangle$. 
An estimate of the gluon condensate is thus obtained by evaluating the
action density in euclidean spacetime $\left\langle
F_{\mu \nu }^{2}\right\rangle =\left\langle \mathcal{L}_{YM}^{euc}\right%
\rangle$ \cite{Diakonov:1995ea,Shifman:1978bx}.

In the framework of the effective theory for thermalized, deconfining SU(2) Yang-Mills dynamics 
we evaluate the action density by virtue of Eq.\,(\ref{lageff}). Thus the average $\left\langle
F_{\mu \nu }^{2}\right\rangle $ corresponds to%
\begin{equation}
\left\langle F_{\mu \nu }^{2}\right\rangle =4\left\langle \mathcal{L}%
_{YM}^{euc}\right\rangle \propto 4\left\langle \mathcal{L}_{{\tiny \mbox{eff}%
}}^{u.g.}\right\rangle =4\rho _{gs}=16\pi \Lambda ^{3}T\,,
\end{equation}%
where a proportionality between the average $\left\langle \mathcal{L}%
_{YM}^{euc}\right\rangle $ over fundamental fields and the average $%
\left\langle \mathcal{L}_{{\tiny \mbox{eff}}}^{u.g.}\right\rangle $ evaluated in the effective theory 
with coarse-grained fields holds \cite{GH2007}. We notice that the ground-state energy
density $\rho_{gs}$ is responsible for the emergence of a nonvanishing
thermal average $\left\langle F_{\mu \nu }^{2}\right\rangle$. 

More precisely, the gluon condensate should be defined as a 
renormalization-group invariant object. This holds for  $\left\langle \frac{%
\beta (g)}{2g}F_{\mu \nu }^{a}F^{a,\mu \nu }\right\rangle _{T}$ where $\beta(g)$ is the full 
beta-function for the fundamental coupling $g$, compare with \cite{GH2007}. 
By virtue of the trace anomaly we can 
evaluate this quantity as
\begin{equation}
\left\langle \frac{\beta (g)}{2g}F_{\mu \nu }^{a}F^{a,\mu \nu }\right\rangle
_{T}=\rho -3p\,.  \label{betaeq}
\end{equation}%
In \cite{lg} we have shown that within the effective theory a linear
growth $\rho-3p=6\rho _{gs}=24\pi \Lambda ^{3}T$ for $T\gg T_{c}$ follows. 
Such a linear 
growth has been found by lattice simulations \cite{millerrev,miller,Langfeld} 
and also in an analytical approach \cite{zw}. In the latter a
momentum-dependent, universal modification of the 
dispersion relation for propagating, fundamental gluon fields, 
motivated by the reduction of the physical state space a la Gribov, is
introduced. While this is interesting a direct comparison of 
both approaches beyond the observation of a linear growth of the trace
anomaly would need a coarse-graining over the modified gluon propagation 
of \cite{zw}.

\section{Conclusions}

In this article we have derived the potential $V(\left\vert \phi \right\vert
^{2})=\Lambda ^{6}/\left\vert \phi \right\vert ^{2}$ for an inert, adjoint
scalar field $\phi$ by solely assuming its origin to be a spatial average
over noninteracting, BPS saturated topological field configurations in the
underlying theory: SU(2) Yang-Mills thermodynamics being in its deconfining
phase. That is, no detailed microscopic information on these configurations
other than their stability and BPS saturation is needed to derive the
potential for the effective field $\phi$. The conceptually interesting
implication of our present work is that the Yang-Mills scale $\Lambda$
emerges as a constant of integration: $\Lambda$'s existence needs not be
assumed as in \cite{Hofmann}. For our presentation to be selfcontained we
have repeated the derivation of the effective action, involving the field $%
\phi$ as a background, for the coarse-grained, topologically trivial
fluctuations. We also have pointed out that the (linear) temperature
dependence of the gluon condensate agrees with that found in lattice
simulations.

There is a host of applications of SU(2) Yang-Mills thermodynamics in
particle physics \cite{SHG} and cosmology \cite{GH}.

\end{document}